\documentclass[journal=jctcce,manuscript=article]{achemso}
\usepackage{graphicx}
\usepackage{amsmath}
\usepackage{physics}
\usepackage{braket}
\usepackage{amssymb}
\usepackage{mathtools}
\usepackage{multirow}
\usepackage[colorlinks,citecolor=blue,urlcolor=black]{hyperref}

\def\ben{\begin{equation}}
\def\een{\end{equation}}
\def\mr{\mathbf{r}}
\def\mg{\mathbf{G}}
\def\mq{\mathbf{q}}
\def\mk{\mathbf{k}}
\def\tmg{\tilde{\mathbf{G}}}
\def\tmq{\tilde{\mathbf{q}}}

\title{Accelerating $GW$-Based Energy Level Alignment Calculations for Molecule-Metal Interfaces Using a Substrate Screening Approach}

\author{Zhen-Fei Liu}
\altaffiliation{Present Address: Department of Chemistry, Wayne State University, Detroit, MI 48202, USA}
\affiliation{Molecular Foundry, Lawrence Berkeley National Laboratory, Berkeley, California 94720, USA}
\alsoaffiliation{Department of Physics, University of California, Berkeley, California 94720, USA}
\author{Felipe H. da Jornada}
\affiliation{Department of Physics, University of California, Berkeley, California 94720, USA}
\alsoaffiliation{Materials Sciences Division, Lawrence Berkeley National Laboratory, Berkeley, California 94720, USA}
\author{Steven G. Louie}
\affiliation{Department of Physics, University of California, Berkeley, California 94720, USA}
\alsoaffiliation{Materials Sciences Division, Lawrence Berkeley National Laboratory, Berkeley, California 94720, USA}
\author{Jeffrey B. Neaton}
\affiliation{Molecular Foundry, Lawrence Berkeley National Laboratory, Berkeley, California 94720, USA}
\alsoaffiliation{Department of Physics, University of California, Berkeley, California 94720, USA}
\alsoaffiliation{Kavli Energy Nanosciences Institute at Berkeley, Berkeley, California 94720, USA}
\email{jbneaton@lbl.gov}

\date{\today}

\begin{document}

\begin{abstract}
The physics of electronic energy level alignment at interfaces formed between molecules and metals can in general be accurately captured by the \emph{ab initio} $GW$ approach. However, the computational cost of such $GW$ calculations for typical interfaces is significant, given their large system size and chemical complexity. In the past, approximate self-energy corrections, such as those constructed from image-charge models together with gas-phase molecular level corrections, have been used to compute level alignment with good accuracy. However, these approaches often neglect dynamical effects of the polarizability and require the definition of an image plane. In this work, we propose a new approximation to enable more efficient $GW$-quality calculations of interfaces, where we greatly simplify the calculation of the non-interacting polarizability, a primary bottleneck for large heterogeneous systems. This is achieved by first computing the non-interacting polarizability of each individual component of the interface, e.g., the molecule and the metal, without the use of large supercells; and then using folding and spatial truncation techniques to efficiently combine these quantities. Overall this approach significantly reduces the computational cost for conventional $GW$ calculations of level alignment without sacrificing the accuracy. Moreover, this approach captures both dynamical and nonlocal polarization effects without the need to invoke a classical image-charge expression or to define an image plane. We demonstrate our approach by considering a model system of benzene at relatively low coverage on aluminum (111) surface. Although developed for such interfaces, the method can be readily extended to other heterogeneous interfaces. 
\end{abstract}

\section{Introduction}

Accurate understanding and determination of electronic energy level alignment at molecule-metal interfaces, i.e., the relative position between molecular frontier resonance orbital energies and the Fermi level of the metal, is critical in understanding the interfacial electronic structure and charge dynamics \cite{KUW13}. As an example, the level alignment of a molecular junction is directly related to its low-bias conductance and current-voltage characteristics \cite{DKCV10}. More heuristically, the level alignment at molecule-metal interfaces can be thought of as the energy barrier for charge transfer across the interface \cite{HRKH98}. Moreover, a quantitative description of level alignment is a prerequisite in understanding and predicting functional properties of a wide range of interfaces, especially those related to heterogeneous catalysis \cite{MMIZ13}, charge transport \cite{N14}, solar energy harvesting \cite{UGNT14}, and other energy conversion processes in the nanoscale \cite{HZSC18}. Electronic energy level alignment is a physical observable, and can be determined experimentally by, e.g., scanning tunneling spectroscopy or photoemission spectroscopy \cite{I99}. However, different binding geometries of the molecule on the metal substrate often lead to different level alignment and charge dynamics \cite{DKKR03,QKSC09}, and thus care must be taken in interpreting experimental data. For example, for molecular junctions, level alignment and conductance often result from an ensemble average of various binding geometries \cite{QVCL07}.

For the purpose of elucidating clear structure-property relationships for molecule-metal interfaces, first-principles electronic structure calculations play an indispensable role, providing information complementary to experiments, through calculations of individual, well-defined geometries. From a formal theory point of view, the levels of interest at interfaces are quasiparticle energy levels, as they are associated with particle-like charged excitations in an interacting system. Additionally, the molecular resonance orbitals of interest are typically a few eV's away from the Fermi level of the metal and are thus not frontier orbitals of the combined system. Therefore, there is no guarantee that these non-frontier orbital energies calculated from Kohn-Sham (KS) density functional theory (DFT) are accurate relative to experiments \cite{P85b}. In fact, common local and semi-local approximations of the exchange-correlation functional usually underestimate the level alignment by about 1 eV or more \cite{NHL06}, leading to incorrect predictions of interfacial charge dynamics, such as significant overestimations of low-bias conductances of molecular junctions \cite{QVCL07}. A possible practical approach within the DFT framework for more accurate level alignment is to employ hybrid functionals in the generalized Kohn-Sham (GKS) scheme (Ref. \citenum{LERK17} is such an example); however, the results depend on the functional, and a wise choice of parameters can be non-trivial for many cases. Although accurate excited-state quantum chemistry approaches, such as those based on the equation of motion \cite{K08}, are well-developed for finite systems such as molecules, having a continuum of electronic states from the metal is indispensable in computing properties of extended interfaces and far from standard in quantum chemistry methods. Additionally, their unfavorable scaling prohibits routine calculations of systems involving hundreds of atoms.

All of the above considerations have contributed to the rise in popularity and success of many-body perturbation theory (MBPT) approaches based on the interacting Green's function formalism, a formally rigorous theoretical framework for computing quasiparticle energies. Within this formulation, the $GW$ approximation \cite{H65,HL86} to the electronic self-energy $\Sigma$ is often employed, where $G$ is the single-particle Green's function and $W$ is the screened Coulomb interaction, and where the perturbative expansion is performed on top of a mean-field calculation, typically KS DFT. This scheme, known as the \textit{ab initio} $GW$ approach, has been shown to be successful in computing quasiparticle excitations for a wide range of materials, including molecules \cite{SCSR15}, bulk solids \cite{HL86,MC13}, and interfaces \cite{KMH14,CTQ17}. Still, one significant bottleneck in traditional $GW$ calculations is in the construction, within the random-phase approximation (RPA), of the non-interacting polarizability matrix $\chi^0(\mathbf{r},\mathbf{r}'; \omega)$, where $\mathbf{r}$ and $\mathbf{r'}$ are real-space coordinates and $\omega$ is a frequency, which has a formal scaling of ${\cal O}(N^4)$ \cite{BerkeleyGW} with $N$ being the system size. Along with other factors, it significantly hampers routine $GW$ calculations of large interface systems. In this work, we focus on overcoming this bottleneck, and propose a practical and simplified solution, specifically for weakly coupled heterogeneous interfaces.

The success of the \textit{ab initio} $GW$ approach for accurately predicting level alignment at interfaces can be attributed to its ability in capturing surface polarization, which is a long-range correlation effect \cite{NHL06,TR09}. More specifically, for a molecule adsorbed on a metal surface, the single-particle excitations in the molecule induce a density response not only within the molecule, but also within the metal surface; the density response within the metal surface in turn modifies the effective strength of the electron-electron interaction within the molecule. This substrate screening \cite{I72} due to the metal surface then alters the charged excitation energies in the molecular subspace compared to the case of an isolated molecule, reducing the energy required to tunnel either an electron or a hole from the surface to the frontier orbitals of the adsorbate, thus reducing (or renormalizing) the  HOMO-LUMO gap (HOMO denotes the highest occupied molecular orbital, and LUMO the lowest unoccupied molecular orbital) \cite{NHL06}. The response of the electrons in the metal to an external charge (located in the vacuum region), or image-charge effect \cite{LK73}, gives rise to an image-charge-like potential far away from the metal surface \cite{I71,AB85,LN93}. This long-range many-body effect is completely missing in the orbital energies in the KS formulation of DFT, since the KS electronic states only see local potentials. Simplified $GW$-based self-energy correction accounting for substrate screening have been developed \cite{NHL06,ELNK15}, where an image-charge model is often used to account for the metallic substrate. These methods, often referred to as DFT+$\Sigma$, can be accurate in the weak-coupling limit \cite{NHL06,QVCL07,TDQB11}; but they require the definition of an image-plane \cite{ELNK15} and they neglect dynamical effects of the electrode polarizability. Beyond DFT+$\Sigma$, recently there have been a number of studies focusing on efficient substrate screening \cite{LVL13,UBSJ14,BUJQ15,QJL17,ALT15,T17,TPV17,CB18,XCQ19}, especially at interfaces involving two-dimensional materials. The essential idea is that the non-interacting RPA polarizability of the interface can be well approximated by the sum of the non-interacting RPA polarizabilities of each individual components \cite{LVL13,UBSJ14,QJL17,XCQ19}, with the latter further approximated using a variety of approaches. Efficiently computing the polarizability, or equivalently the dielectric function that is responsible for many-body effects, has been an active area of research \cite{R10,LVL13,XD15,DDNM18}.

In this work, we develop a simplified approach for efficient calculations of the non-interacting RPA polarizability for molecule-metal interface systems. This approach greatly reduces the computational cost of $GW$ calculations of weakly coupled heterogeneous interfaces without sacrificing the accuracy, enabling calculations on computer clusters of moderate size. Our approach takes advantage of the fact that the non-interacting RPA polarizability of a weakly coupled molecule-metal interface can be well approximated as being additive, i.e., as the sum of the non-interacting RPA polarizability of the metal substrate and that of the periodic molecular layer. This is inspired by the concepts in Ref. \citenum{UBSJ14}, where the non-interacting RPA polarizability of a MoSe$_2$-bilayer graphene interface is approximated by the sum of the non-interacting RPA polarizability of MoSe$_2$ and that of a bilayer graphene. However, unlike Ref. \citenum{UBSJ14}, where in-plane local field effect is neglected for the substrate (i.e., the effect of the substrate screening is assumed to be the same if the substrate is displaced along any of its extended directions), we do not rely on this approximation here, and instead compute it efficiently and without additional approximations by \emph{folding} this quantity in reciprocal space from a much smaller unit cell. Additionally, we introduce a real-space  scheme for efficiently calculating the non-interacting RPA polarizability of the periodic molecular layer, avoiding calculations using large vacuum regions and substantially reducing the computational cost. The non-interacting RPA polarizabilities of the two individual components are then combined in the original supercell of the interface, followed by the standard $GW$ approach for evaluating the self-energy corrections. We show that our new method yields almost identical level alignments as the direct $GW$ calculations of weakly coupled molecule-metal interfaces, while significantly reducing the computational cost, as we demonstrate for benzene adsorbed on Al(111) surface. Finally, even though we develop our method in the context of molecule-metal interfaces, we expect that it is straightforward to extend this approach to other heterogeneous interfaces, such as those formed between molecules and semiconductors and atomically thin systems supported by substrates.

The structure of this article is as follows. In Sec. \ref{sec:theory}, we discuss in detail the approach we develop, showing the main assumption, i.e., additivity of the non-interacting RPA polarizability, and the ideas in reducing the computational cost for each individual components. In Sec. \ref{sec:results}, we use one example, benzene adsorbed on Al(111), to show numerically that the result is almost identical to direct $GW$ calculations of level alignment. We also demonstrate the efficiency of our approach by comparing computational resources needed to those required for a direct $GW$ calculation. Sec. \ref{sec:discuss} is devoted to discussions of our approach and possible future directions. Then we conclude in Sec. \ref{sec:conclude}.

\section{Theoretical Approach}
\label{sec:theory}



A major ingredient and bottleneck of \emph{ab initio} $GW$ calculations in the standard formulation \cite{HL86} is the non-interacting RPA polarizability $\chi^0$:
\begin{equation}
\chi^0(\mathbf{r},\mathbf{r}';\omega) = \sum_{i}^{\rm occ.} \sum_{a}^{\rm vir.} \frac{\phi_i(\mathbf{r})\phi_a^*(\mathbf{r})\phi_a(\mathbf{r}')\phi_i^*(\mathbf{r}')}{\omega+\epsilon_i-\epsilon_a}.
\label{chi0}
\end{equation}
In this equation, $i$ runs over all occupied KS orbitals (denoted by $\phi$) and $a$ runs over all virtual (unoccupied) KS orbitals of the system. $\epsilon$'s are the KS eigenvalues. A direct evaluation of this equation, or the equivalent one in reciprocal space, leads to a formal scaling of ${\cal O}(N^4)$, where $N$ is the number of basis functions used to express $\phi_i(\mathbf{r})$ and $\phi_a(\mathbf{r})$.

\begin{figure}[htbp]
\begin{center}
\includegraphics[width=3.5in]{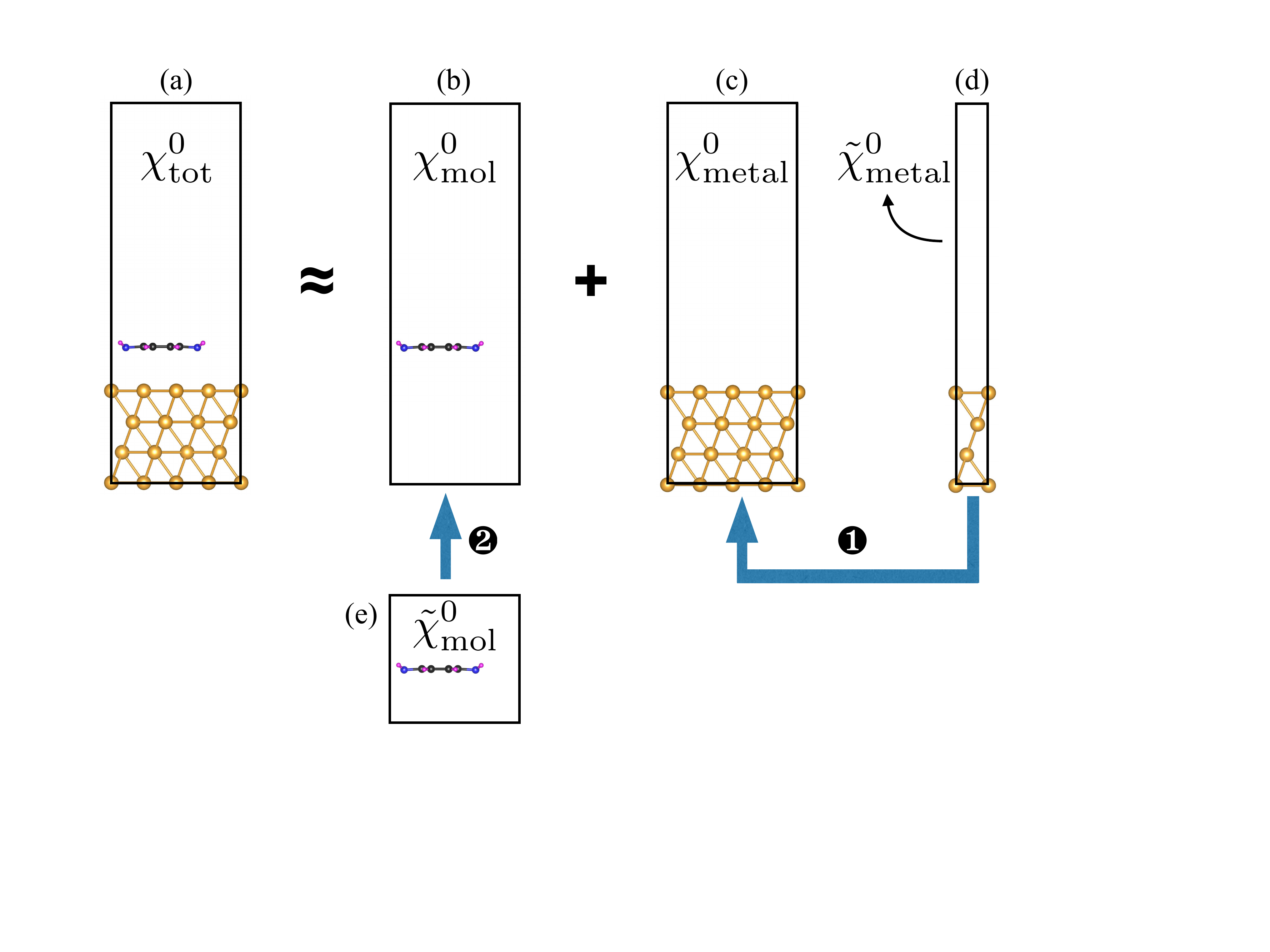}
\caption{Schematic representation of the main approximations in this work. Instead of computing $\chi^0$ for system (a) directly, we propose to compute $\tilde{\chi}^0$ for systems (d) and (e), and approximate $\chi^0$ of (a) by the sum of $\chi^0$ of (b) and (c), which can be in turn calculated efficiently from $\tilde{\chi}^0$ of (e) and (d), respectively. Black boxes denote supercells and periodic boundary conditions for all directions.}
\label{f:cartoon}
\end{center}
\end{figure}

The evaluation of $\chi^0$ can be simplified for a combined system consisting two subsystems -- a molecular layer and a metal substrate -- if there is no significant hybridization. In this limit, the occupied states $\phi_{i_\mathrm{mol/metal}}$ can be partitioned so that they belong either to the molecular layer or to the metal substrate, while the virtual states can be approximated as either localized resonances in either one of the subsystems $\phi_{a_\mathrm{mol/metal}}$, or a free electronic state $\phi_{a_\mathrm{free}}$. Hence,  cross terms such as $\phi_{i_\mathrm{mol}}(\mathbf{r})\phi_{a_\mathrm{metal}}^{*}(\mathbf{r})$ which have small overlap can be dropped, and we approximate the sum in Eq. \eqref{chi0} as
\begin{equation}
\sum_{i}^{\rm occ.} \sum_{a}^{\rm vir.}
\rightarrow
\sum_{i\in \rm{mol}}^{\rm occ.}\sum_{a\in \rm{mol}/\rm{free}}^{\rm vir.} + \sum_{i\in \rm{metal}}^{\rm occ.}\sum_{a\in \rm{metal}/\rm{free}}^{\rm vir.}.
\end{equation}
In this limit (and other scenarios discussed in Ref. \citenum{XCQ19}), $\chi^0$ is separable and additive, i.e.,
\begin{equation}
\chi^0_{\rm tot}(\mathbf{r},\mathbf{r}') \approx \chi^0_{\rm mol}(\mathbf{r},\mathbf{r}') + \chi^0_{\rm metal}(\mathbf{r},\mathbf{r}').
\label{add}
\end{equation}
This equation plays the central role in this work and is shown schematically in Fig. \ref{f:cartoon}. For simplicity, we use the Hybertsen-Louie generalized plasmon pole (GPP) model \cite{HL86} for calculations in this article; however, the approach can be readily generalized to fully frequency-dependent treatment of $\chi^0$, and so we drop the frequency dependence in all equations below. 

In panel (a) of Fig. \ref{f:cartoon}, we show the side view of the supercell of a generic molecule-metal interface, which consists of a molecule physisorbed on a metallic slab substrate. Periodic boundary conditions are used for all three directions, though the long-range Coulomb interaction is truncated \cite{I06} along the direction perpendicular to the substrate, which we denote by $z$. Despite the truncation, a large vacuum region in the supercell along $z$ is needed to avoid spurious interactions between neighboring cells. 

In Eq. \eqref{add}, $\chi^0_{\rm tot}$ is the non-interacting RPA polarizability of the combined molecule-metal interface system, which is represented in panel (a). Similarly, the non-interacting RPA polarizability of the periodic molecular layer only, $\chi^0_{\rm mol}$, is represented in panel (b), while that of the metal slab alone, $\chi^0_{\rm metal}$, is in panel (c). We emphasize that we use the same supercells when computing $\chi^0_{\rm tot}$, $\chi^0_{\rm mol}$, and $\chi^0_{\rm metal}$, as represented in panels (a), (b), and (c). Because the number of unoccupied orbitals needed in the sum in Eq. \eqref{chi0} to converge the result scales as the volume of the supercell, merely invoking Eq. \eqref{add} does not lead to a drastic speedup of the calculation if the supercells are kept the same.

We introduce two procedures to accelarate the computation of the non-interacting RPA polarizability. On the right of Fig. \ref{f:cartoon}, we schematically represent the first approximation, by which we obtain $\chi^0_{\rm metal}$ in the supercell geometry [Fig. \ref{f:cartoon}(c)] by \emph{folding} from a calculation performed in the primitive unit cell of the metal slab $\tilde{\chi}^0_{\rm metal}$ [Fig. \ref{f:cartoon}(d)], as we detail in the following section. Here,  $\tilde{\chi}^0_{\rm metal}$ is the non-interacting RPA polarizability of the periodic metal substrate in a primitive unit cell. This primitive unit cell for the metal slab may be much smaller along the extended directions, but we keep the lattice constant in the out-of-plane $z$ direction the same as in the supercell geometry. Because the number of atoms and plane-wave components required to express the polarizability matrix are then significantly reduced, this procedure accelarates computing the polarizability of the metal substrate. We note that we neglect geometry optimization for the metal surface in this work, and assume all metal atoms are in their bulk positions. We can then map $\tilde{\chi}^0_{\rm metal}$ of the metal slab calculated in its primitive unit cell [Fig. \ref{f:cartoon}(d)] to that calculated in the supercell geometry,  $\chi^0_{\rm metal}$  in Fig. \ref{f:cartoon}(c), without loss of information. Because $\tilde{\chi}^0_{\rm metal}$ is evaluated in a larger reciprocal cell than ${\chi}^0_{\rm metal}$, this procedure is equivalent to \textit{folding} $\tilde{\chi}^0_{\rm metal}$ into a smaller region of the Brillouin zone. This procedure, denoted as process 1 with a blue arrow in Fig. \ref{f:cartoon}, will be elaborated in Sec. \ref{sec:foldmetal} below.

The second procedure we develop is to employ a smaller cell for the subproblem of the periodic molecular layer [Fig. \ref{f:cartoon}(e)]. This supercell, shown in Fig. \ref{f:cartoon}(e), has the same dimensions in-plane, along the extended directions, but may be considerably smaller along the $z$ direction compared to the supercell of the combined system [Fig. \ref{f:cartoon}(b)]. As a result, it is more computationally efficient to compute the non-interacting RPA polarizability of the molecular layer in this smaller supercell, $\tilde{\chi}^0_{\rm mol}$, than that of the original supercell, $\chi^0_{\rm mol}$. The conversion between $\tilde{\chi}^0_{\rm mol}$ and $\chi^0_{\rm mol}$ is denoted as process 2 with a blue arrow in Fig. \ref{f:cartoon}, and it is achieved by a spatial truncation of $\chi^0_{\rm mol}$ in real space. This will be discussed in detail in Sec. \ref{sec:foldmol} below.

\subsection{Folding of the Polarizability for the Metal Substrate}
\label{sec:foldmetal}
In this section, we describe the approach that allows us to efficiently compute the non-interacting RPA polarizability of the metal slab in its primitive unit cell and map it to a larger supercell.

We define our supercell for the combined system in terms of three lattice vectors, $\vb{a}_1$, $\vb{a}_2$, and $\vb{a}_3$. The system has translational symmetry in the $xy$ plane spanned by $\vb{a}_1$ and $\vb{a}_2$, and is confined along the perpendicular direction $\vb{a}_3$, which is parallel to the $z$ axis, $\hat{a}_3=\hat{z}$. The metallic substrate is approximated as a finite slab with $N_\mathrm{uc}$ atoms in its primitive unit cell and defined by the lattice vectors $\tilde{\vb{a}}_1\equiv\vb{a}_1/N_1$, $\tilde{\vb{a}}_2\equiv\vb{a}_2/N_2$, and $\tilde{\vb{a}}_3\equiv\vb{a}_3$, where $N_i$ are integers. Intrinsic properties of the substrate can then either be computed in this primitive unit cell [Fig. \ref{f:cartoon}(d)], say, with a $M_1N_1 \times M_2N_2$ $\mk$-point mesh, or equivalently in a supercell that is compatible with the relevant molecular coverage [Fig. \ref{f:cartoon}(c)], with $N_\mathrm{sc}=N_\mathrm{uc}N_1 N_2$ metal atoms and lattice vectors $\vb{a}_1$, $\vb{a}_2$, and $\vb{a}_3$, and with a $M_1 \times M_2$ $\mk$-point mesh. The former is much more computationally efficient, since the computational cost of the dielectric response scales at most quadratically with the number of the $\mk$ points, while it displays a $\mathcal{O}(N_\mathrm{at}^3)$ to $\mathcal{O}(N_\mathrm{at}^4)$ computational cost with respect to the number of atoms $N_\mathrm{at}$ in the supercell.

We denote the non-interacting RPA polarizability of the metallic substrate in the supercell [Fig. \ref{f:cartoon}(c)] by $\chi^0_{\rm metal}$; in reciprocal space, it is written as $\chi^0_{{\rm metal}\,\mathbf{G},\mathbf{G}'}(\mathbf{q})\equiv \chi^0_{\rm metal}(\mathbf{q}+\mathbf{G},\mathbf{q}+\mathbf{G}')$, where $\mathbf{G}$ and $\mg'$ are reciprocal lattice vectors, and $\mq$ is a vector in the first Brillouin zone. We also denote the non-interacting RPA polarizability of the metallic slab in its primitive unit cell [Fig. \ref{f:cartoon}(d)] by $\tilde{\chi}^0_{\rm metal}$; in reciprocal space, we write it as $\tilde{\chi}^0_{{\rm metal}\,\tmg,\tmg'}(\tmq)\equiv \tilde{\chi}^0_{\rm metal}(\tmq+\tmg,\tmq+\tmg')$. Note that $\chi^0_{\rm metal}$ and $\tilde{\chi}^0_{\rm metal}$ describe the same quantity for the same physical excitation, i.e., ${\chi}^0_{\rm metal}(\mq+\mg,\mq+\mg')=\tilde{\chi}^0_{\rm metal}(\tmq+\tmg,\tmq+\tmg')$ when $\mq + \mg = \tmq + \tmg$ and $\mq + \mg' = \tmq + \tmg'$ in Cartesian coordinates. However, even though any combination of vector sums $\mq+\mg$ in the supercell can be written as  $\tmq+\tmg$ in the primitive unit cell, the individual components of $\mq$ and $\tmq$ or $\mg$ and $\tmg$ may differ since, in general, the $\tmg$ in the primitive unit cell are a subset of the $\mg$ in the supercell. Correspondingly, there are more $\mg$ vectors in the supercell up to a given energy cutoff than there are $\tmg$ vectors in the primitive cell up to same cutoff. A similar analysis holds for the $\mathbf{q}$ points in the Brillouin zone: the $\mathbf{q}$-point mesh for the primitive unit cell is $N_1 \times N_2$ times denser than that for the supercell, and so we conclude that there are $N_1 \times N_2$ smaller matrices $\tilde{\chi}^0_{{\rm metal}\,\tmg,\tmg'}(\tmq)$ in the primitive unit cell which, together, hold the same amount of physical information as a single larger matrix $\chi^0_{{\rm metal}\,\mathbf{G},\mathbf{G}'}(\mathbf{q})$ in the supercell. For a specific $\mq$ point in the Brillouin zone of the supercell, the $N_1 \times N_2$ $\tmq$ points that fold to $\mq$ satisfy: $\tmq = \mq + n_1 \mathbf{b}_1 + n_2 \mathbf{b}_2$, where $n_1 = 0, 1, 2, \cdots, N_1-1$, and $n_2 = 0,1,2, \cdots, N_2-1$, where $\mathbf{b}_1$ and $\mathbf{b}_2$ are the two reciprocal lattice unit vectors of the supercell associated with the lattice vectors $\vb{a}_1$ and $\vb{a}_2$.

\begin{figure}[htbp]
\begin{center}
\includegraphics[width=3.in]{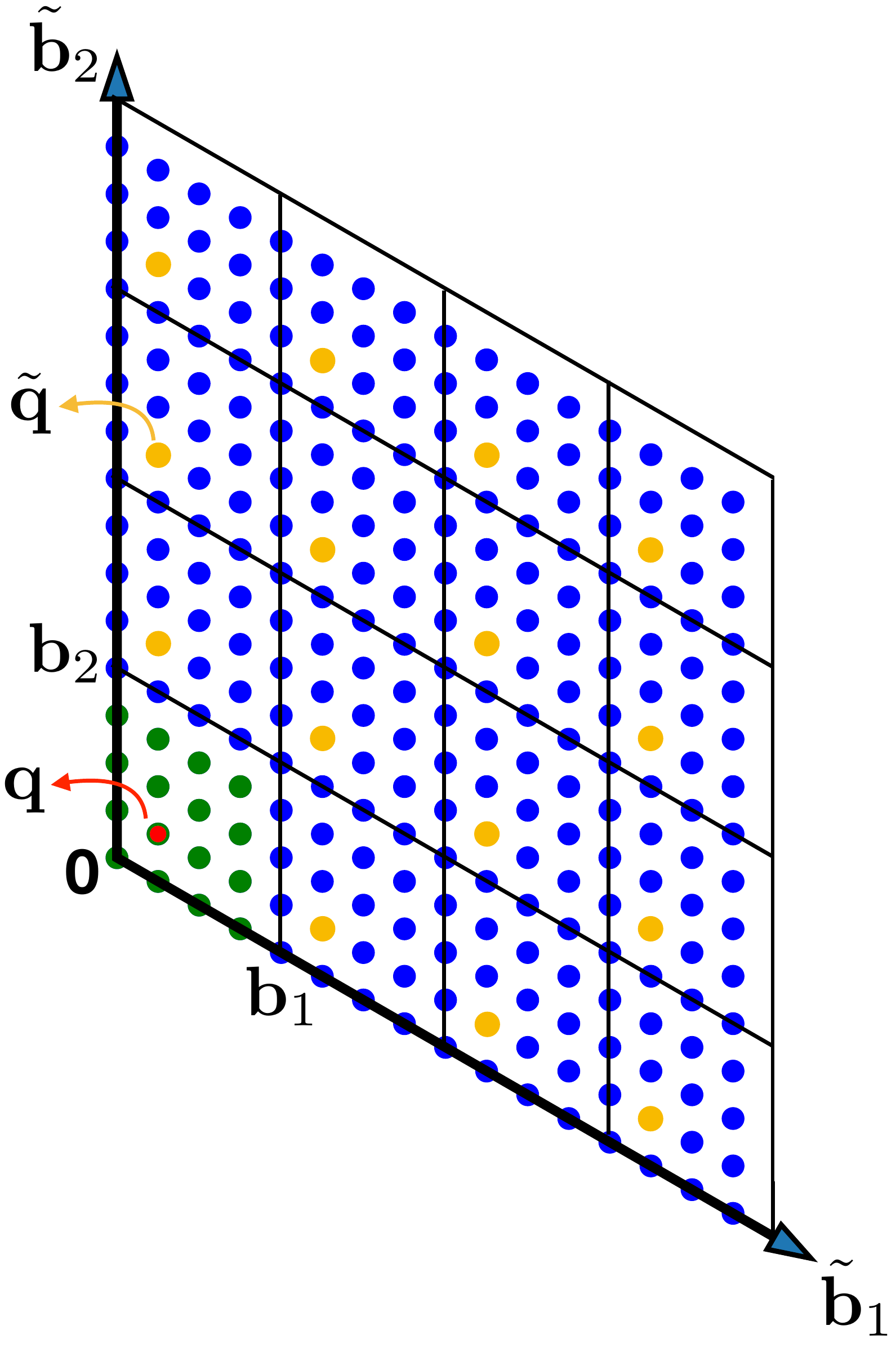}
\caption{The folding of the non-interacting RPA polarizability of the metal substrate in the reciprocal space. $\mathbf{b}_1$ and $\mathbf{b}_2$ are the two reciprocal lattice unit vectors of the supercell. $\tilde{\mathbf{b}}_1$ and $\tilde{\mathbf{b}}_2$ are the two reciprocal lattice unit vectors of the primitive unit cell. The $\mq$-point sampling for the supercell is shown as green dots and the $\mq$-point sampling for the primitive unit cell is shown as blue dots. The red dot denotes an arbitrary $\mq$ point in the supercell, and it is folded from all the $\tmq$ points in the primitive unit cell that are shown as the yellow dots.}
\label{f:foldmetal}
\end{center}
\end{figure}

The folding procedure is shown schematically in Fig. \ref{f:foldmetal}, where $\mathbf{b}_1$ and $\mathbf{b}_2$ are the two reciprocal lattice vectors of the supercell, $\tilde{\mathbf{b}}_1$ and $\tilde{\mathbf{b}}_2$ are the two reciprocal lattice vectors of the primitive unit cell, and they are related as $\tilde{\mathbf{b}}_1=N_1 \mathbf{b}_1$ and $\tilde{\mathbf{b}}_2=N_2 \mathbf{b}_2$. The $\mathbf{q}$-point sampling for the supercell (smaller region in reciprocal space) is represented by green dots, and the $\mathbf{q}$-point sampling for the primitive unit cell (larger region in reciprocal space) is represented by blue dots. In this figure, we display the actual sampling for the system that we study in this work, namely, benzene on Al(111), which we discuss in Sec. \ref{sec:results}. In this example,  there are $N_\mathrm{uc}=4$ Al atoms in the primitive unit cell of the metallic slab (1 atom in each layer). The supercell is defined by $N_1=N_2=4$, which contains $N_\mathrm{sc}=64$ Al atoms (16 atoms in each layer), and it is sampled with a $M_1\times M_2=4\times 4$ $\mq$-point grid. Accordingly, the calculation of the primitive unit cell is performed with a $N_1M_1\times N_2M_2=16\times 16$ $\mq$-point grid. For a given $\mq$ point in the supercell, say, one denoted by the red dot in Fig. \ref{f:foldmetal}, we need to combine information from $N_1 \times N_2$ $\tmq$ points in the primitive unit cell that are shown as the yellow dots in Fig. \ref{f:foldmetal}.

\subsection{Construction of the Spatially Truncated Polarizability Matrix for the Periodic Molecular Layer}
\label{sec:foldmol}
The observation that $\chi^0_{\rm mol}(\mr,\mr')$ is only non-negligible when both $\mr$ and $\mr'$ are near the molecule suggests that one can use a smaller cell to compute this quantity, avoiding high computational cost associated with large vacuum in plane-wave based $GW$ calculations. In Fig. \ref{f:cartoon}(e), the length $\tilde{L}_z$ of the supercell along the $z$ direction is smaller than the length $L_z$ in the original supercell shown in panel (b), though they share the same dimensions along the extended directions, as we discussed above. The $\mathbf{q}$-point sampling along the extended directions used in the smaller cell containing the molecule [Fig. \ref{f:cartoon}(e)] should be equivalent to that of the original supercell [Fig. \ref{f:cartoon}(b)]. Additionally, the length $\tilde{L}_z$ should not be too small as to introduce spurious interactions between neighboring cells along $z$ and large numerical errors. Here, we find that $\tilde{L}_z/L_z$ of the order of  1/2 or 1/3 significantly reduces the computational cost without incurring noticeable numerical errors. For simplicity, we refer to the original supercell corresponding to Fig. \ref{f:cartoon}(b) as the ``large-$z$'' cell, and the smaller cell for the molecular-layer calculation, shown in Fig. \ref{f:cartoon}(e), as the ``small-$z$ cell''. We also keep the convention of putting a tilde over all quantities calculated in the smaller cell. We then approximate $\chi^0_{\rm mol}(\mr,\mr')$ in the large-$z$ cell in the following way: set $\chi^0_{\rm mol}(\mr,\mr')=\tilde{\chi}^0_{\rm mol}(\tilde{\mr},\tilde{\mr}')$ in the region defined by the small-$z$ cell, and set all other elements of $\chi^0_{\rm mol}(\mr,\mr')$ to zero. This procedure then constitutes a spatial truncation of $\chi^0_{\rm mol}(\mr,\mr')$ in real space.

Due to the size of the $\tilde{\chi}^0_{{\rm mol}\,\tmg,\tmg'}(\tmq)$ matrix for typical interfaces, it is neither computationally advantageous nor necessary to perform three-dimensional Fourier transforms on its two indices $\tmg$ and $\tmg'$ to obtain $\tilde{\chi}^0_{\rm mol}(\tilde{\mr},\tilde{\mr}')$ in real space for subsequent mapping to $\chi^0_{\rm mol}(\mr,\mr')$. Rather, we take advantage of the fact that the small-$z$ cell and the large-$z$ cell share the same dimensions along the two extended directions. Hence they can be labeled by the same quantum numbers $(\mq;G_1,G_2)$ along the extended directions, and the Fourier transforms only need to be carried out along the confined direction $z$.

Thus, starting from the non-interacting RPA polarizability computed in the small-$z$ cell, for each combination of wave vector and reciprocal lattice vectors $(\mq;G_1,G_2;G_1',G_2')$, we perform Fourier transforms only along the $z$ direction and obtain $\tilde{\chi}^0_{\rm mol}(\tilde{z},\tilde{z}')$. The $(\mq;G_1,G_2;G_1',G_2')$ indices are understood implicitly here, since they are the same for both the large-$z$ cell and the small-$z$ cell. After we compute $\tilde{\chi}^0_{\rm mol}(\tilde{z},\tilde{z}')$, we simply insert this matrix into the larger matrix $\chi^0_{\rm mol}(z,z')$ and set the elements outside of the domain of $\tilde{\chi}^0_{\rm mol}(\tilde{z},\tilde{z}')$ to zero. This is physically equivalent to performing a spatial truncation of $\chi^0_{\rm mol}(z,z')$ in the large-$z$ cell. Finally, we Fourier transform ${\chi}^0_{\rm mol}(z,z')$ back to obtain  $\chi^0_{\rm mol}(G_z,G_z')$. We note that the Fourier transforms are performed with a discrete fast Fourier transform (FFT) algorithm, so, importantly, they are performed on discretized grids (commonly referred to as FFT grids) that are compatible with the different lattice constants $L_z$ and $\tilde{L}_z$.

\begin{figure}[htbp]
\begin{center}
\includegraphics[width=3.5in]{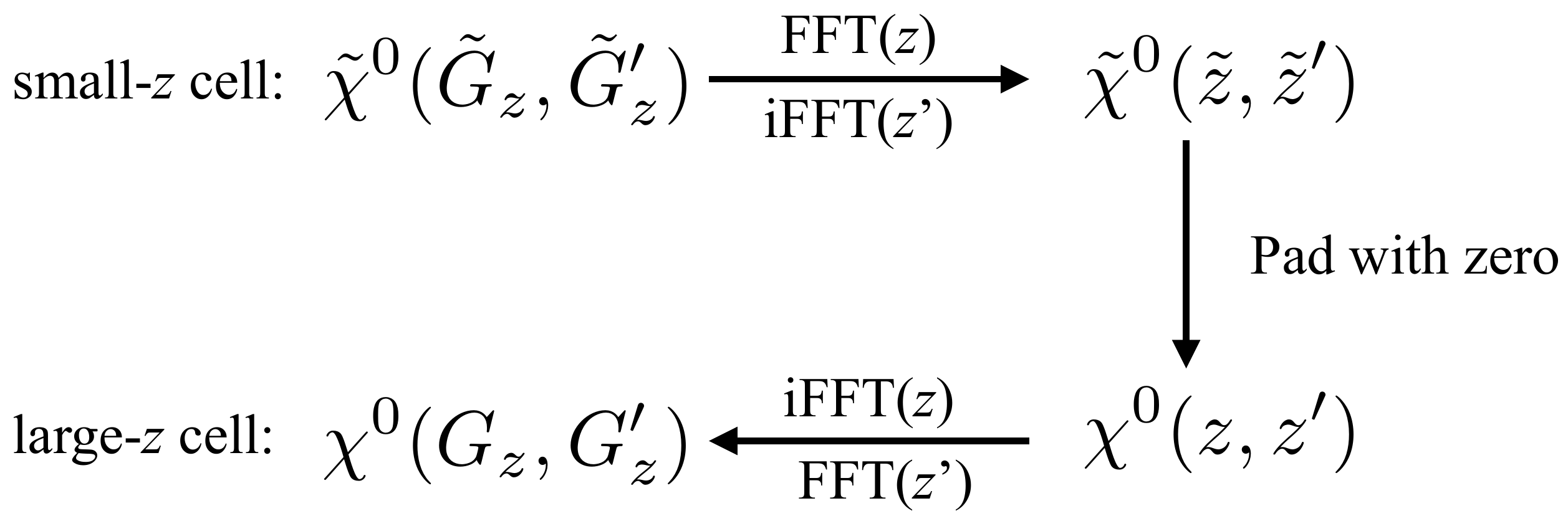}
\caption{The workflow showing the mapping of the non-interacting RPA polarizability of the periodic molecular layer in the small-$z$ cell to that of the large-$z$ cell. This procedure is performed for each combination of $(\mq;G_1,G_2;G_1',G_2')$ indices. iFFT denotes inverse FFT.}
\label{f:foldmol}
\end{center}
\end{figure}

Overall, this procedure is repeated for each composite index  $(\mq;G_1,G_2;G_1',G_2')$ and hence can be parallelized trivially. The workflow for this procedure is described in Fig. \ref{f:foldmol}.

\subsection{Workflow}

We briefly summarize the workflow for the method proposed in this work. The goal is to calculate $GW$ self-energy corrections to levels of the combined molecule-metal interface system that correspond to molecular frontier resonance orbitals. The non-interacting RPA polarizability of the combined system is approximated as in Eq. \eqref{add}, and the self-energy correction is then calculated in the standard manner within the \textit{ab initio} $GW$ approach. We assume the geometry is properly optimized and only mention the electronic structure calculations below. The workflow has 10 steps, as follows,
\begin{enumerate}
\item
Perform a DFT calculation of the combined system, and identify the levels of the combined system that correspond to molecular frontier resonance orbitals. This can be done, e.g., by expanding the HOMO of the molecular layer using all the orbitals of the combined system as the basis and then identifying the largest projection.
\item
Choose a primitive unit cell for the metal substrate, which can be repeated along the $\vb{a}_1$ and $\vb{a}_2$ directions to form the substrate of the combined system. This is shown in Fig. \ref{f:cartoon}(d). The length of the supercell along $z$-direction should be the same as that of the combined interface system. Perform a self-consistent DFT calculation of this system.
\item
Calculate the non-interacting RPA polarizability $\tilde{\chi}^0_{{\rm metal}\,\tmg,\tmg'}(\tmq)$ for the primitive unit cell of the metal slab. 
\item
Fold the above quantity to the supercell, ${\chi}^0_{{\rm metal}\,\mg,\mg'}(\mq)$. This step is schematically shown as process 1 with a blue arrow in Fig. \ref{f:cartoon}.
\item
Choose a small-$z$ cell for the molecular layer, with $\vb{a}_1$ and $\vb{a}_2$ the same as in the large-$z$ cell. This corresponds to the cell in Fig. \ref{f:cartoon}(e). Perform a self-consistent DFT calculation of this system. Note that it is physically a molecular layer instead of a single molecule, so the same k-mesh is needed as the original interface system.
\item
Calculate the non-interacting RPA polarizability $\tilde{\chi}^0_{{\rm mol}\, \tmg,\tmg'}(\tmq)$ for the small-$z$ cell for the molecular layer. 
\item
Map the above quantity to the large-$z$ cell, ${\chi}^0_{{\rm mol}\,\mg,\mg'}(\mq)$. This step is shown as process 2 with a blue arrow in Fig. \ref{f:cartoon}.
\item
Add the quantities calculated in Step 4 and Step 7. The matrix size should be the same and is equivalent to the  polarizability matrix that would have been calculated directly for the interface. We denote the result by $\chi^0_{\rm tot}$.
\item
Obtain the inverse dielectric matrix of the combined system, $\epsilon^{-1}_{\mg,\mg'}(\mq)=\left[1-v(\mq)\chi^0_{{\rm tot}\,\mg,\mg'}(\mq)\right]^{-1}$.
\item
Use this inverse dielectric matrix and the wavefunctions of the combined system to calculate self-energy corrections to the levels that correspond to molecular frontier resonances (determined in Step 1) as in the standard $GW$ approach. Note that a large number of unoccupied bands for the combined system is still required for the summation of the screened exchange \cite{BerkeleyGW}, due to the fact that the final quasiparticle energy levels converge slowly with respect to the number of states summed up in the evaluation of the self-energy operator \cite{SXZC10}.
\end{enumerate}
For the above steps listed, Steps 1, 9, and 10 are also present in a direct $GW$ calculations of the interface. Steps 2-8 are proposed in this work, where Steps 2-4 are calculations for the substrate only, and Steps 5-7 are for the periodic molecular layer only.

\section{Results}
\label{sec:results}

\begin{figure}[htbp]
\begin{center}
\includegraphics[width=3.5in]{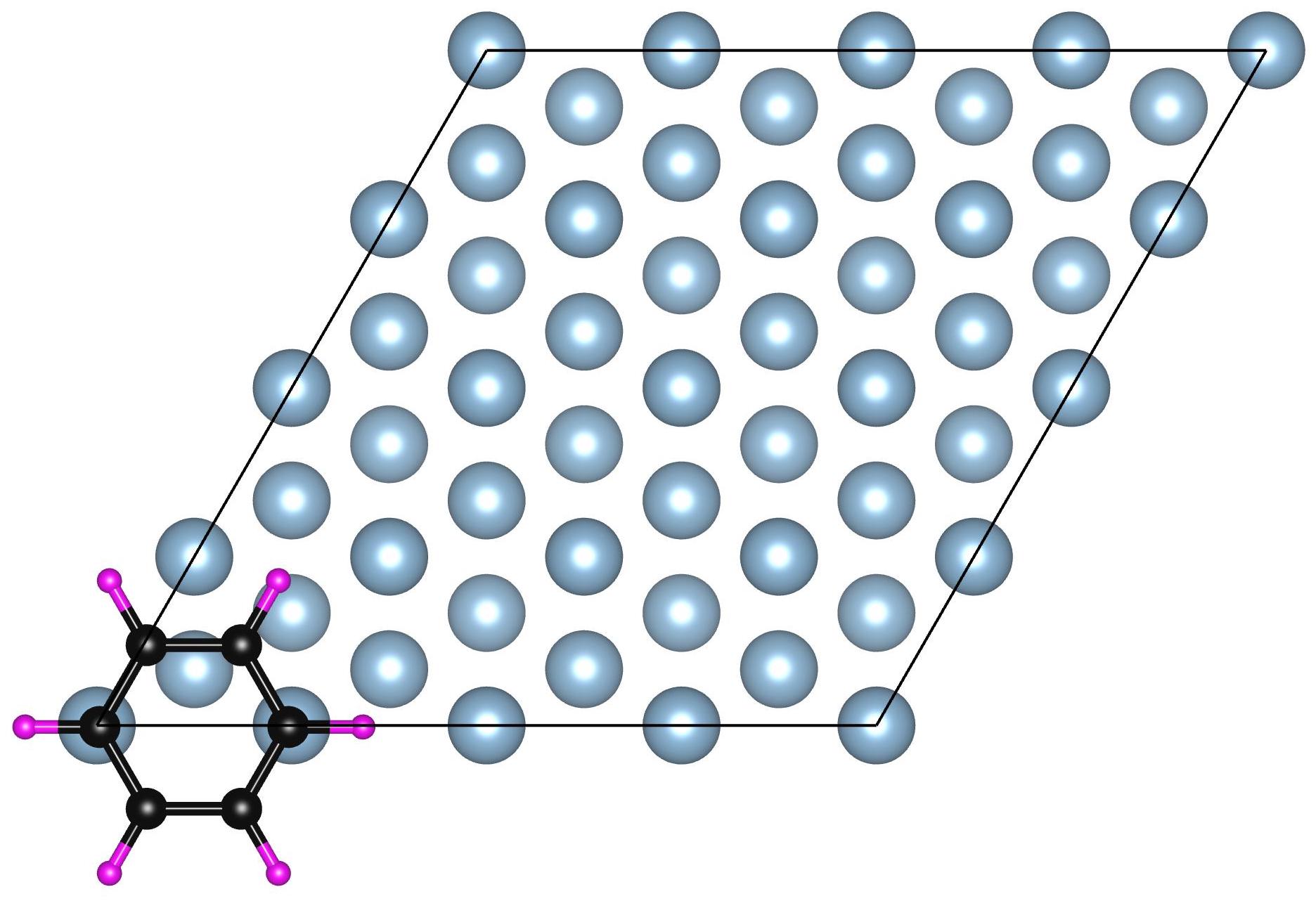}
\caption{The test-study system used in this work, benzene molecule sitting at 3.24 \AA~above Al(111) surface. A top view is shown. We also consider another case of 4.24 \AA~for the adsorption height. Color codes for the atoms: light blue - aluminum; black - carbon; magenta - hydrogen.}
\label{f:bzal}
\end{center}
\end{figure}

We present a test study using a previously well-studied \cite{LERK17}, experimentally relevant \cite{DMBN00} weakly coupled system: benzene physisorbed flat on Al(111) surface, where we consider one benzene molecule per $4\times4$ surface Al atoms, as shown in Fig. \ref{f:bzal}. The geometry we use is the following: relaxed benzene molecule sitting on a hollow site 3.24 \AA~above Al(111) surface \cite{LERK17} (we also consider the case of 4.24 \AA~in Sec. \ref{sec:accuracy} below). In its primitive unit cell, the Al metal slab contains 4 atoms with 1 atom on each layer, and the corresponding supercell is constructed with $4\times 4$ repetition of the primitive unit cell, with 64 Al atoms (16 atoms on each layer). The supercell of the interface has length $L_z=33$ \AA~along $z$, and the corresponding small-$z$ cell of the molecular layer is $\tilde{L}_z=11$ \AA. The system is shown in Fig. \ref{f:bzal}. A $4\times 4 \times 1$ $\mathbf{q}$-point mesh is used for the interface system and the small-$z$ molecular layer calculations, and a $16\times 16 \times 1$ $\mathbf{q}$-point mesh is used in the unit cell calculations of the metal substrate. The Cartesian coordinates of these $\mathbf{q}$ points are shown in Fig. \ref{f:foldmetal}. All DFT calculations are performed with the \texttt{Quantum ESPRESSO} package \cite{QE2017} using the PBE functional \cite{PBE96} and an energy cutoff of 50 Ry. All $GW$ calculations are performed with the \texttt{BerkeleyGW} package \cite{BerkeleyGW}. An energy cutoff of 5 Ry is used for all polarizability calculations, which corresponds to 4976 bands in the interface supercell (311 bands in the primitive unit cell of the metal slab and 1659 bands in the small-$z$ cell of the molecular layer). Metal screening and slab truncation of Coulomb interaction \cite{I06} are used for both dielectric function (\texttt{epsilon}) and self-energy (\texttt{sigma}) steps in \texttt{BerkeleyGW}. We discuss two aspects of the method proposed here: accuracy, by comparing results from different approximations; and efficiency, by comparing computing resources used in the calculations.

\subsection{Verification of the Approach}
\label{sec:accuracy}

Table \ref{tab1} shows the accuracy of our approach. The left column shows the method and the level of approximation. We list the $GW$ level alignment for both HOMO ($E_F-E_{\rm HOMO}^{GW}$) and LUMO ($E_{\rm LUMO}^{GW}-E_F$). At the PBE level, the HOMO resonance of the molecule is located at 3.17 eV below the Fermi level ($E_F$) of the interface, and the LUMO resonance of the molecule is located at 2.01 eV above $E_F$. We note that although these calculations are reasonably well converged, it is not our goal here to fully test the convergence with respect to energy cutoff, $\mk$-point sampling, and number of bands used in $GW$ calculations for this system. Rather, we focus on the accuracy of our approximation at a fixed level of convergence, such that all parameters are scaled accordingly when we change the size of the supercell.

\begin{table}[htp]
\caption{Comparison of different levels of approximation for the calculation of molecular resonance level alignments with respect to Fermi level of the interface, for benzene molecule sitting flat at 3.24 \AA~above Al(111) surface. Here, $\Delta E$ is the difference between the $GW$ quasiparticle energy and the KS orbital excitation energy from PBE. All values are in eV.}
\begin{center}
\begin{tabular}{c|cc|cc}
\hline\hline
Method & $\Delta E_\mathrm{HOMO}$ & $E_F-E_{\rm HOMO}^{GW}$ & $\Delta E_\mathrm{LUMO}$ & $E_{\rm LUMO}^{GW}-E_F$ \\
\hline\hline
Direct $GW$ of the interface & 0.72 & 3.89 & 0.36 & 2.38 \\
\hline
Use $\chi^0_{\rm tot} \approx \chi^0_{\rm metal} + \chi^0_{\rm mol}$, with & \multirow{2}{*}{0.88} & \multirow{2}{*}{4.05} & \multirow{2}{*}{0.41} & \multirow{2}{*}{2.43} \\
$\chi^0_{\rm metal/mol}$ from direct calculations & & & & \\
\hline
$\chi^0_{\rm metal}$ from folding, and & \multirow{2}{*}{0.88} & \multirow{2}{*}{4.05} & \multirow{2}{*}{0.41} & \multirow{2}{*}{2.43} \\
$\chi^0_{\rm mol}$ from real-space mapping & & & & \\
\hline\hline
\end{tabular}
\end{center}
\label{tab1}
\end{table}

We note that the difference between the direct $GW$ and the result employing the approximation $\chi^0_{\rm tot} \approx \chi^0_{\rm metal} + \chi^0_{\rm mol}$ originates from the fact that there is hybridization between molecular orbitals and metal orbitals. In fact, the largest overlap \cite{note1} $\left|\braket{\phi_{\rm tot}|\phi_{\rm mol}}\right|^2$ at the $\Gamma$ point for HOMO is about 0.79 and that for LUMO is about 0.54. Therefore, Eq. \eqref{add} is an approximation and does not hold exactly, meaning that the polarizability for the combined system is not strictly additive. To verify that this is the case, we tested the case where we artificially move the benzene molecule further away by 1 \AA~from the surface and place it flat at 4.24 \AA~above Al(111) surface. For this system, the largest $\left|\braket{\phi_{\rm tot}|\phi_{\rm mol}}\right|^2$ at the $\Gamma$ point for HOMO is about 0.85 and that for LUMO is about 0.82. Here, $\chi^0_{\rm tot} \approx \chi^0_{\rm metal} + \chi^0_{\rm mol}$ is an even better approximation and, correspondingly, the $\Delta E_{\rm HOMO}$ is only 0.07 eV different from that in direct $GW$ calculations. Nevertheless, even for the system we consider in Table \ref{tab1}, the difference in $\Delta E_{\rm HOMO}$ is only 0.16 eV and the difference in $\Delta E_{\rm LUMO}$ is only 0.05 eV.

We also note that the folding of $\chi^0_{\rm metal}$ is in principle exact. However, in our calculations, we found that typically one gets a $0.1\%$ to $0.5\%$ relative error in this procedure, defined as $\|\chi^0_\mathrm{metal, folded}-\chi^0_\mathrm{metal, direct}\|_{\rm F}/\|\chi^0_\mathrm{metal, folded}+\chi^0_\mathrm{metal, direct}\|_{\rm F}$, where $\|X\|_{\rm F}$  denotes the Frobenius norm of matrix $X$, and $\chi^0_\mathrm{metal, folded}$ and $\chi^0_\mathrm{metal, direct}$ denote the non-interacting polarizability of the metal slab in the supercell geometry computed with the folding procedure and via the direct approach, respectively. We suspect that this small numerical error is related to the slightly different way the energy cutoff over reciprocal lattice vectors is enforced on different $\mq$ points due to the fact that the $\mg$ vectors are discrete, and hence this error should diminish with cutoff and completely vanish in the limit of an infinite plane-wave cutoff. 

The $\mq=0$ point of metal requires special treatment. Within \texttt{BerkeleyGW} \cite{BerkeleyGW}, a special $\mq_0$ (small but nonzero) is used. In the folding procedure we propose in this work, the calculation of the unit cell also requires a special $\tmq_0$ whose Cartesian coordinates must match that of the $\mq_0$ in the supercell. The $\chi^0_{\rm metal}(\mq_0)$ matrix of the supercell is then folded from this $\tilde{\chi}^0_{\rm metal}(\tmq_0)$ and other regular $\tmq$ points that fold to $\mq=0$ point (not $\mq_0$) of the supercell. This procedure also yields an even smaller error in $\chi^0_{\rm metal}(\mq_0)$, depending on the number of bands summed in the calculation of $\tilde{\chi}^0(\tmq_0)$. In practice, only a small number of bands are needed \cite{BerkeleyGW} for this calculation. Alternatively, since the $\chi^0(\mq_0)$ for the combined system is much more efficient to compute than $\chi^0(\mq)$ because only a small number of bands are needed, it can also be calculated directly for the interface with an affordable computational cost, without invoking any folding from the primitive unit cell.

\subsection{Efficiency}

The computational efficiency of our approach is demonstrated in Table \ref{tab2} by listing CPU hours used and the total memory required in each step. We note that the actual performance and timing highly depend on the architecture of the computer cluster used, though the qualitative trend should remain the same. Our test calculations reported in this section are performed using the computer cluster ``Etna'' (Intel Xeon E5-2670 v3, base frequency 2.30 GHz) at the Molecular Foundry at Lawrence Berkeley National Laboratory. 

\begin{table}[htp]
\caption{Comparison of computing resources for the direct calculation of non-interacting RPA polarizability and the approach proposed in this work. CPU hours is defined as the number of processors used multiplied by the number of wall-time hours needed for the calculation. Total memory is defined as the minimum amount of memory required for each processor in $\chi^0$ calculations multiplied by the total number of processors used, measured in gigabytes (GB). Memory is not a bottleneck for non-$GW$ calculations so we do not list those numbers. In this table, nscf stands for non-self-consistent calculations used to generate Kohn-Sham bands for the summation in $\chi^0$ required by \texttt{BerkeleyGW}.}
\begin{center}
\begin{tabular}{c|c|c|c|c}
\hline\hline
\multirow{2}{*}{Method} & \multirow{2}{*}{Breakdown} & Subtotal & Total & Total Memory \\
& & CPU hours & CPU hours & Required (GB) \\
\hline
Direct $GW$ & - & 3550 & 3550 & 1257\\
\hline
\multirow{7}{*}{This work} & Metal nscf in unit cell & 30 & \multirow{7}{*}{1016} & - \\
& Metal $\chi^0$ in unit cell & 339 & & 61\\
& Metal $\chi^0$ folding & 1 & & -\\
& Molecular layer nscf in small-$z$ cell & 194 & & -\\
& Molecular layer $\chi^0$ in small-$z$ cell & 83 & & 110\\
& Molecular layer $\chi^0$ mapping & 369 & & -\\
\hline\hline
\end{tabular}
\end{center}
\label{tab2}
\end{table}

We first discuss the CPU resources needed for $\chi^0$ calculation. We can see that the mapping of $\chi^0$ for the periodic molecular layer contributes to a large amount of computing time in our proposed work. This is because one needs to loop through all the $(G_1,G_2)$ pairs, all the $(G_1',G_2')$ pairs, and all the $\mq$ points; and in each of these loops, one must perform four Fourier transforms: two with dimension $\tilde{N}_z$ and with cost $\mathcal{O}(\tilde{N}_z \log \tilde{N}_z)$, and another two with dimension $N_z$ and with cost $\mathcal{O}(N_z \log N_z)$, where $\tilde{N}_z$ ($N_z$) is the number of FFT grids along the $z$ direction in the small-$z$ (large-$z$) cell. For the system considered here, our approach already takes as little as  30\% of the computing time of the direct $GW$ calculation of the interface, which we expect to be an upper bound. More importantly, this speedup should grow for larger systems, as the cost of the folding procedure becomes negligible compared to the cost of computing the polarizability matrix for larger unit cells. In fact, we have done another benchmark calculation using benzene on graphite(0001) surface (same structure as that in Ref. \citenum{NHL06}, results not shown), where our proposed method takes less than 20\% of the computing resource needed for the direct $GW$ calculation. 

More crucially, we note that the direct calculation of $\chi^0$ for the combined interface system requires large memory in addition to the large CPU resources. It is this large memory requirement that typically renders $GW$ calculations for large systems infeasible on small computer clusters. With the approach proposed here, we not only reduce the amount of CPU resources needed, but also reduce significantly the memory needed for $GW$ calculations, as the evaluation of $\chi^0$ is now performed in much smaller supercells. Our approach will therefore enable $GW$ calculations for large interface systems on computer clusters of much smaller size.

\subsection{Accelarating Self-Energy Calculations}

For all results shown above, we perform the standard one-shot $GW$ self-energy calculations using the GPP after the $\chi^0_{\rm tot}$ is computed with our approach. For weakly coupled interfaces, we can further reduce the computational cost of the self-energy calculations by assuming the difference in self-energy corrections between the GPP and static Coulomb-hole-screened-exchange (COHSEX) \cite{JDSC14} is the same for the combined interface system and the molecular layer in the small-$z$ cell. Similar ideas based on the physics presented in Ref. \citenum{NHL06} were also used in Refs. \citenum{UBSJ14} and \citenum{CTQ17} to speed up $GW$ calculations of interface systems. Specifically, we propose:
\begin{equation}
\begin{split}
\braket{\phi_{\rm tot}|\Sigma^{\rm GPP}_{\rm tot}|\phi_{\rm tot}}\approx & \braket{\phi_{\rm tot}|\Sigma^{\rm COHSEX}_{\rm tot}|\phi_{\rm tot}} \\
& + \braket{\phi_{\rm mol}|\Sigma^{\rm GPP}_{\rm mol}|\phi_{\rm mol}} \\
& -\braket{\phi_{\rm mol}|\Sigma^{\rm COHSEX}_{\rm mol}|\phi_{\rm mol}}.\\
\end{split}
\label{sediff}
\end{equation}
In this equation, $\phi_{\rm tot}$ is the KS orbital of the combined interface system that represents the molecular resonance, $\phi_{\rm mol}$ is the frontier orbital of the molecular layer, calculated in the small-$z$ cell. The second and third terms on the right hand side of \eqref{sediff} can be calculated in the small-$z$ cell following the standard $GW$ self-energy calculations, which are computationally efficient. The first term on the right hand side, although calculated in the large-$z$ cell, only involves occupied states due to the use of static COHSEX \cite{HL86} and therefore is also computationally efficient. Therefore, the left hand side of Eq. \eqref{sediff}, responsible for $GW$ energy level alignment, can be calculated efficiently. Using Eq. \eqref{sediff} leads to \emph{additional} savings of computing resources on top of what is shown in Table \ref{tab2}, because we avoid GPP calculations for the entire interface, which requires many empty bands to converge $\Sigma^{\rm GPP}_{\rm tot}$. Calculating KS bands is an ${\cal O}(N^3)$ process, and by using Eq. \eqref{sediff}, we only need to generate empty bands for the small-$z$ cell.

Eq. \eqref{sediff} works well for weakly coupled molecular resonances, i.e., when the orbital is relatively unchanged from gas phase to interface, usually satisfied for physisorption. For the system of benzene sitting at 3.24 \AA~above Al(111) surface, the HOMO resonance is weakly coupled. The difference between the two sides of Eq. \eqref{sediff} is only 0.04 eV. However, Eq. \eqref{sediff} applies less well to LUMO of this system, because $\left|\braket{\phi_{\rm tot}|\phi_{\rm mol}}\right|^2$ is only 0.54 as we discussed in Sec. \ref{sec:accuracy}. For the system of benzene sitting at 4.24 \AA~above Al(111) surface, both HOMO and LUMO are weakly coupled, as we discussed in Sec. \ref{sec:accuracy}, and the difference between the two sides of Eq. \eqref{sediff} is about 0.08 eV and 0.18 eV for HOMO and LUMO, respectively.
 
\section{Discussion}
\label{sec:discuss}
The calculation of non-interacting RPA polarizability $\chi^0$ has a formal scaling of ${\cal O}(N^4)$ in a standard plane-wave approach, and within \texttt{BerkeleyGW}, prior to the calculation of $\chi^0$, a large number of bands need to be generated for the system under study to express the Green's function in its spectral representation. The generation of bands has a formal scaling of ${\cal O}(N^3)$, and the number of bands scales with the volume of the supercell. Our approach here partitions a calculation in a large cell involving many bands into two calculations in much smaller cells involving far fewer bands: one is the primitive unit cell of the metal, which is $N_1 \times N_2$ times smaller than the original system, and the other is the small-$z$ cell of the periodic molecular layer, which is a fraction of the size  of the original cell. We note that our approach can also be combined with $GW$ formalisms based on localized basis (such as Ref. \citenum{molgw}, but note that one still needs periodic boundary conditions for the molecular layer), or occupied orbitals only \cite{GCL10,GG15}, and it would be of interest to apply this approach within those frameworks. 

We expect that our approach can greatly reduce the computational cost for systems requiring a large energy cutoff, having large number of valence electrons, or having a large supercell. Examples include systems with $d$-orbitals, where it is known that semi-core electrons need to be treated explicitly \cite{RKP95,LICL02,MOD02} and therefore the energy cutoff is high and the number of electrons is large. We showed that, with the additivity of the non-interacting RPA polarizability as simple as Eq. \eqref{add}, we achieve very accurate level alignment compared to a direct $GW$ calculation. The error that we obtain is only on the order of 0.1 eV, which is useful in making quantitative predictions for experimental observables.

Besides the efficiency and the accuracy, our approach captures substrate screening, or ``image-charge effects'', without the need of defining an image plane, and treats the periodic nature of the molecular layer explicitly. This is a step forward from the perturbative, static DFT+$\Sigma$ approach \cite{NHL06,ELNK15} in three aspects. First, we eliminate the need for the definition of an image plane, which sometimes could be non-trivial, especially for semiconductor surfaces \cite{ZHCZ16}. Although we present our work in the context of molecule-metal interfaces here, we expect that it can be extended to semiconductor interfaces in a straightforward manner. Second, the DFT+$\Sigma$ approach does not treat the periodic molecular monolayers rigorously, as the classical image-charge interaction is only between the substrate and the adsorbate, so it does not capture the long-range Coulomb interaction within the periodic molecular layer \cite{LERK17} and therefore does not capture intra-layer orbital renormalization \cite{TS11}. Our approach presented here treats the molecular layer with periodic boundary conditions rigorously within the \textit{ab initio} $GW$ approach and could therefore capture these effects naturally. This is particularly useful for studying high-coverage surfaces \cite{TDWL10}, or the covalent organic frameworks \cite{RFDP17} on metal surfaces, where in both cases the periodic nature of the adsorbate layer is critical and cannot be overlooked. Third, the DFT+$\Sigma$ approach is static, in the sense that it neglects dynamical (frequency-dependent) effects in the non-interacting RPA polarizability or dielectric function. On the other hand, our approach captures these effects naturally, either with a plasmon-pole model\cite{HL86}, as we use in this work, or in principle also by explicitly computing the frequency-dependent dielectric function.

The method proposed here will be less effective if the coupling between the molecular layer and the substrate is too strong, such as the cases of covalently bound interfaces \cite{TJPL95} where a chemical bond is broken when the molecule is adsorbed on the substrate. This is because the additivity of the non-interacting RPA polarizability, Eq. \eqref{add}, will break down in those cases. It would nonetheless be interesting to perform a series of tests for the intermediate cases, such as those with pinned molecular levels at $E_F$ \cite{ZFSP10}, or those involving strong substrate-enhanced dispersion \cite{ULRK14}. It would be more interesting to generalize Eq. \eqref{add}, adding extra necessary corrections, for strongly hybridized or covalently bound interfaces. Moreover, the computational cost will be further reduced if the self-energy can be evaluated effectively within the supercell of the molecular layer only, instead of within the supercell containing the combined interface system as we do in the present work. This will require effective embedding of the molecule within the dielectric environment of the substrate. Studies along these lines are currently underway. 

\section{Conclusions}
\label{sec:conclude}

In this work, we develop an efficient $GW$-based approach for calculating the level alignments at molecule-metal interfaces, greatly reducing the computational cost of such calculations without significantly sacrificing the accuracy. It is based on the approximation that the non-interacting RPA polarizability of the combined interface system can be expressed as the sum of the non-interacting RPA polarizabilities from the two individual subsystems, namely, the metal slab substrate and the molecular layer. The non-interacting RPA polarizability of the substrate is first calculated in the primitive unit cell for the metal slab geometry, followed by a folding scheme in reciprocal space to map the polarizability back to the original supercell of the interface. At the same time, the non-interacting RPA polarizability of the periodic molecular layer is computed in a unit cell that is much smaller along the confined $z$ direction, which is then mapped back in real-space to the original supercell.

Our approach is very accurate for weakly coupled interfaces, reducing the computational cost to only a small fraction of that of a direct $GW$ calculation for typical interface systems. Our approach correctly captures the physics of substrate screening without the need of a full $GW$ calculation or of defining an image plane and, in addition, it also captures dynamical effects of the screening environment. Although our scheme was developed and presented in the context of molecule-metal interfaces, it can be extended to other types of interfaces in a straightforward manner. We expect that this approach will be useful in lowering the cost for routine $GW$ interface calculations, as well as in making these calculations feasible on computer clusters much smaller than those in supercomputing centers.

\begin{acknowledgement}

This work was supported by the U.S. Department of Energy, Office of Basic Energy Sciences, Division of Materials Sciences and Engineering (Theory FWP) under Contract No. DE-AC02-05CH11231, which provided the theoretical formulation and calculations of interfacial screening response. Work performed at the Molecular Foundry was also supported by the Office of Science, Office of Basic Energy Sciences, of the U.S. Department of Energy under the same contract number. This work was additionally supported by the Center for Computational Study of Excited-State Phenomena in Energy Materials (C2SEPEM) at Berkeley Lab, funded by the U.S. Department of Energy under the same contract number, which provided the advanced computer codes. We acknowledge NERSC for computing resources.

\end{acknowledgement}

\bibliography{lit_molscreen}

\begin{mcitethebibliography}{63}
\bibitem[not()]{note1}
Strictly speaking, the absolute value of this quantity depends on the system
  size. Here, we simply use the relative magnitude of this quantity as a
  measure for the strength of coupling between the molecule and the metal
  surface, for two systems in the identical simulation cell, with the same
  energy cutoff, and at the same k-point ($\Gamma$).\relax
\mciteBstWouldAddEndPunctfalse
\mciteSetBstMidEndSepPunct{\mcitedefaultmidpunct}
{}{\mcitedefaultseppunct}\relax
\EndOfBibitem
\end{mcitethebibliography}

\end{document}